\documentclass[%
 reprint,
 twocolumn,
 amsmath,amssymb,
 aps,
prb,
floatfix,
]{revtex4-2}

\usepackage{rotating}
\usepackage{graphicx}
\usepackage{dcolumn}
\usepackage{bm}
\usepackage{amsmath}
\usepackage{amssymb}
\usepackage{subfigure}  
\usepackage{hyperref}
\usepackage{nicefrac}
\usepackage{hyperref}
\hypersetup{colorlinks=true, breaklinks=true, citecolor=blue, linkcolor=blue, menucolor=blue, urlcolor=blue}
\usepackage{ulem}
\usepackage[dvipsnames]{xcolor}

\newcommand{\sqrtthree}{$\left(\sqrt{3}\times \sqrt{3} \right)\textrm{R30}^\circ$}
\newcommand{\PdDTe}{PdTe$_2$}
\newcommand{\PdTeSurf}{TePd$_2$}

\begin{document}


\title{Tellurization of Pd(111): absence of PdTe$_2$ but formation of a TePd$_2$ surface alloy}

\author{Eric Engel}
\author{Alexander Wegerich}
\author{Andreas Raabgrund}
\author{M. Alexander Schneider}%
\email{alexander.schneider@fau.de}
\affiliation{%
 Solid State Physics, Friedrich-Alexander-Universität Erlangen-Nürnberg, Staudtstraße 7, 91058 Erlangen, Germany
}%

\date{March 6, 2024}

\begin{abstract}

In a recent publication [2D Materials, \textbf{8}, 045033 (2021)], it was reported that the growth of a monolayer \PdDTe\ in ultra-high vacuum could be achieved by deposition of tellurium on a palladium (111) crystal surface and subsequent thermal annealing. 
By means of low-energy electron diffraction intensity (LEED-IV) structural analysis, we show that the obtained \sqrtthree\ superstructure is in fact a \PdTeSurf\ surface alloy.  
Attempts to produce a \PdDTe\ layer in ultra-high vacuum by increasing the Te content on the surface were not successful.

\end{abstract}

\maketitle


\section{\label{sec:intro} Introduction}

In the wake of graphene research, there has been a surge in the exploration of two-dimensional materials. 
Transition metal dichalcogenides (TMDCs) of the form MX$_2$ (M = metal, X=S, Se, Te) attracted particular attention.
This interest stems from their distinctive topological properties and potential applications. 
Most of these materials, including \PdDTe\ that is of interest here, can be grown as bulk crystals.
To investigate the physical properties of individual, two-dimensional  sheets of TMDCs, researchers employ the methods of exfoliation and mechanical transfer onto suitable substrates.
This route, however, is not well scalable and not suitable for device fabrication. 
The alternative is an MBE type of growth where a surface reaction between metal and chalcogen is induced on the chosen substrate.
The most straightforward approach involves allowing the chalcogen to react with the transition-metal substrate, resulting in the formation of a TMDC layer.
The method bears similarity to the creation of ultra-thin oxide layers on various metal surfaces.
The success of this approach necessitates a careful evaluation, requiring unambiguous proof of the surface structure.
This is crucial to eliminate alternative possibilities such as chalcogen adsorbate phases or surface chalcogenides \cite{Guan2018,Uenzelmann2020,Kisslinger2020,Kisslinger2021,Geldiyev2023,Kisslinger2023}.

Recently, Liu, Zemlyanov, and Chen reported on the growth of a strained \PdDTe\ layer on Pd(111) in \sqrtthree\ registry \cite{Liu2021}. 
This finding is surprising as the proposed \PdDTe\ layer exhibits strong tensile stain: the lateral lattice parameter of \PdDTe\ is $a_\textrm{\PdDTe}=4.03$\,\AA\ \cite{Soulard2005} while $\sqrt{3}\ a_\textrm{Pd111} = 4.76$\,\AA.
Hence, if this notion is correct, significant forces must act across the interface, which should be attributed to strong Te-Pd bonds between the lower Te layer and the substrate.
Such interfacial bonds are expected to alter the properties of the grown layer significantly towards those of a Pd-Te alloy. 

We reinvestigated the tellurization of Pd(111) and prove here by low-energy electron diffraction intensity analysis (LEED-IV) that the \sqrtthree\ superstructure is in fact a \PdTeSurf\ surface alloy, where one Pd atom per unit cell is exchanged by a Te atom.

\begin{figure*}
	\centering
	\includegraphics[width=0.8\textwidth]{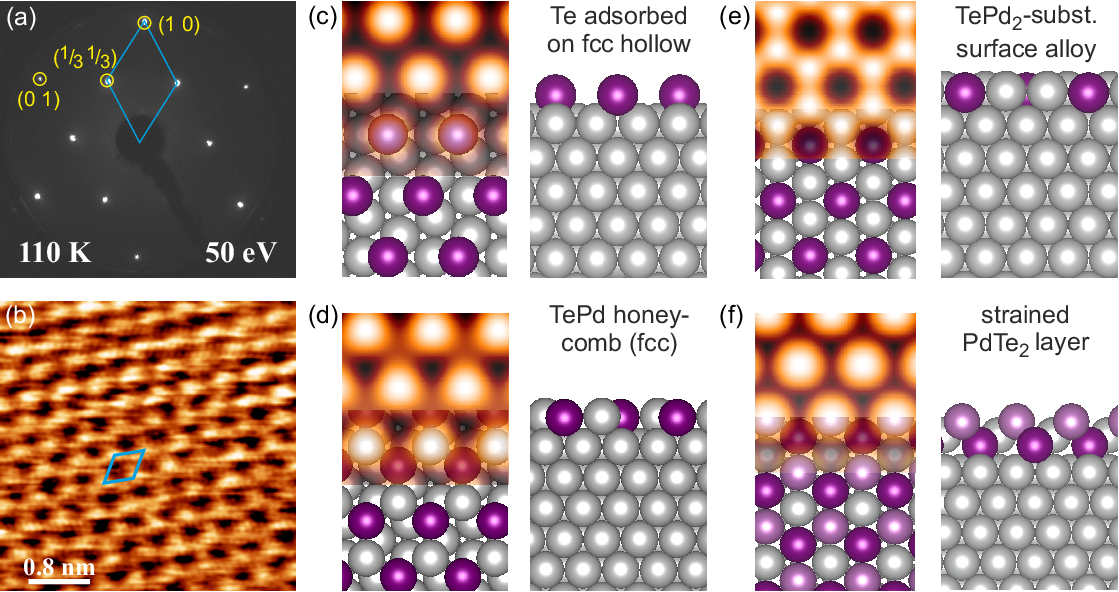}
	\caption{(a) LEED image and (b) STM image showing the \sqrtthree\ superstructure on Pd(111) induced by reaction with 0.33\,ML Te. (Imaging parameters: $U=-0.41$\,V,$I=0.15$\,nA.) (c)-(f): model suggestions and DFT simulated STM images including the suggestion from Ref.\,\cite{Liu2021} in (f). Gray spheres: Pd, purple (dark) spheres: Te.
	}
	\label{Fig1}
\end{figure*}

\section{\label{sec:exp} Methods}

Our experimental and theoretical methods, LEED-IV, scanning tunneling microscopy (STM), and density functional theory (DFT), are described in great detail in our previous publications \cite{Kisslinger2021, Kisslinger2023}.
By comparison with these systems, we were also able to precisely determine the amount of Te evaporated onto the clean Pd(111) crystal.
All experiments were performed under UHV conditions ($p<2\cdot 10^{-10}$\,mbar).
The Pd(111) substrate was cleaned by standard sputtering and annealing cycles.
For structural analysis of the Te-induced Pd(111) \sqrtthree superstructure 0.33 monolayer of Te was evaporated onto the Pd(111) surface held at 90\,K. 
We define one monolayer (ML) to correspond to an adsorbate surface density equal to the atomic density of the underlying substrate (15.3\,nm$^{-2}$).
To induce the surface reaction and the formation of the well-ordered \sqrtthree\ superstructure an annealing temperature of at least 720\,K was necessary.
The structural order could be improved slightly by annealing to temperatures up to 1070\,K. 
Beyond that temperature, decomposition sets in and the \sqrtthree\ reflexes vanish.
We note that Liu et al.\,\cite{Liu2021} also report on the formation of the \sqrtthree\ superstructure after annealing to 470\,$^\circ$C (743\,K).
Liu et al. characterized the Te amount deposited from XPS electron attenuation lengths and stated a Te thickness of approximately 4\,\AA\ which may correspond to a full monolayer of Te. 
From our experiments (see below) we find that any Te in excess of 0.33 ML desorbs from the surface  starting at 540\,K. With these observations, we believe to have prepared the same system as \cite{Liu2021}.\\

In the work presented here, we used the newly developed \mbox{\textsc{ViPErLEED}} package \cite{Riva2021} which provides a sophisticated tool for LEED-IV data acquisition and manages a modified and parallelized \mbox{\textsc{TensErLEED}} code \cite{Blum2001} for full-dynamical calculation of intensity spectra and parameter fitting.
Experimental LEED-IV data were recorded at normal incidence for energies from 50\,eV up to 600\,eV in steps of 0.5 eV and stored for off-line evaluation. The temperature of the substrate during LEED-IV data taking was 110\,K, consequently we used as lattice parameter of the Pd(111)-(1x1) the value of 2.755\,\AA\ determined at that temperature in \cite{Arblaster2012}.
We tested the suggested \PdDTe\ on Pd(111) by Liu et al.\,\cite{Liu2021}, the simple Pd(111)-\sqrtthree-Te adsorbate structure, the Pd(111)-\sqrtthree-PdTe honeycomb structure and the Pd(111)-\sqrtthree-TePd$_2$ substitutional surface alloy against our experimental data. 

The analysis was backed by DFT structural energy relaxations using the VASP package \cite{vasp3} and the PBE-PAW general gradient approximation \cite{PBE}. 
For that, $\left(\sqrt{3}\times \sqrt{3}\right)$-Pd(111) slabs were set up consisting of eight layers of which the three lowest were kept fixed at bulk positions. 
Repeated slabs are separated by at least 1.5\,nm of vacuum.
The STM images were simulated based on the Tersoff-Hamann approximation \cite{Tersoff85}. 

\section{Results}\label{sec:res}

Not surprisingly, our LEED data looks the same as that presented by Liu et al. \cite{Liu2021} (Fig.\,\ref{Fig1}(a)). For the STM image (Fig.\,\ref{Fig1}(b)) we chose data with a slightly different appearance than that in \cite{Liu2021} but depending on tip state and tunneling bias also a regular hexagonal pattern of maxima was observed. 
We aim to convey to the readers that relying solely on STM and DFT is insufficient sometimes for determining a specific surface structure.
While fine details in the DFT may lead to the identification of the correct model (here the substitution of Te in the surface), the dependence of such images on tip state on the experimental side and on parameters of the DFT Tersoff-Hamann simulations can make agreement or disagreement fortuitous.

In Fig.\,\ref{Fig1}(c)-(f) we show the structural models that we tested in our LEED-IV analysis and the corresponding DFT image simulations. 
The simple Te-adsorbate model (c), the TePd-honeycomb (d), and the PdTe$_2$ layer (f) were also tested in hcp stacking sequence.
All models led to converged DFT structures and the DFT simulated images could serve to explain the experimentally observed contrast, although the interpretation of which atoms appear as bright features would be different depending on model. 

\begin{figure}
	\centering
	\includegraphics[width=0.9\columnwidth]{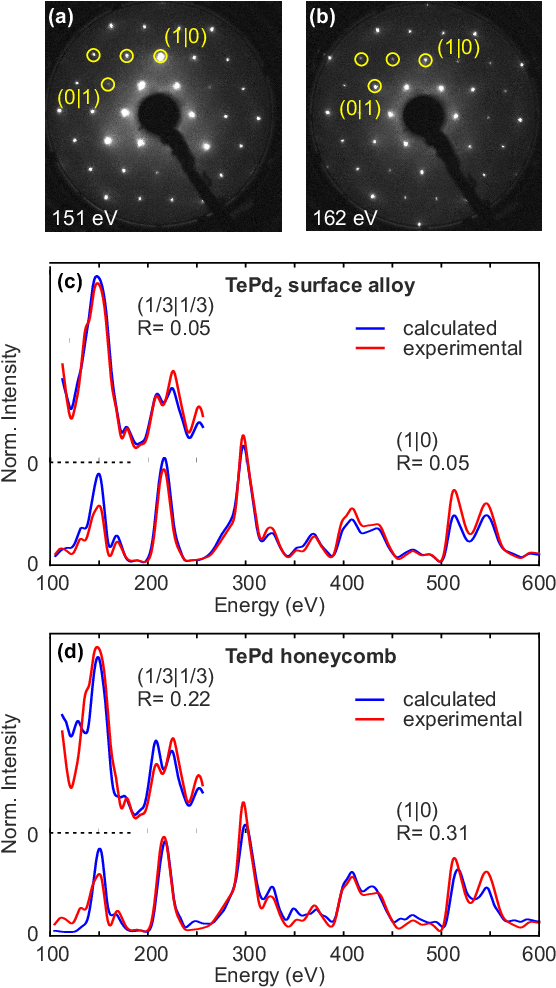}
	\caption{(a) (b) experimental LEED at two further energies demonstrating the idea of  using LEED pattern of a set of particular energies as ``finger prints'' of a particular surface structure (see text). (c) Comparison of the experimental LEED-IV spectra (red) and calculated spectra using for the substitutional model (Fig.\,\ref{Fig1} (e)) after final fine fitting. (d) The same level of fine fitting procedures applied to the honeycomb model introduced in Fig.\,\ref{Fig1}(d). Note that the two models differ by one additional atom only. 
	}
	\label{Fig2}
\end{figure}

By using simple LEED imaging the unit cell is determined only. 
In many cases, this is a redundant information if STM is also available. 
What would be more important is to provide a selection of images at different electron energies that track characteristic intensity variations.
By this it can be verified that upon repetition of an experiment, the same surface phase and not only a phase with the same surface unit cell was prepared.
An example is shown in Figs.\,\ref{Fig2}(a) and (b). 
The system as we prepared it, shows distinctively different intensities of the $(1|0)$ and $(0|1)$ spots at 151\,eV and rather similar intensities at 162\,eV. 
Likewise the $( \nicefrac{2}{3}|\nicefrac{2}{3})$ and $(\nicefrac{1}{3}|\nicefrac{4}{3})$ spots are brighter at 151\,eV than the $(0|1)$ spot and dimmer at 162\,eV.
 
The essence of LEED-IV structural analysis is now to record the intensity of (ideally all) accessible spots, and compare it to the theoretically expected intensity. 
The comparison is governed by the Pendry R-factor \cite{Pendry1980} which is $R=0$ for perfect agreement and $R=1$ completely unrelated spectra.
A level $R < 0.2$ is commonly considered to indicate the correct structural model.

In a first step it was tried to find agreement between experimental and calculated spectra by varying atomic z-coordinates in rather rough steps of \mbox{$\Delta z = 3\,\textrm{pm}$} and $\Delta xy = 5\,\textrm{pm}$. 
For the adsorbate model we find $R =0.57$, for the honeycomb model $R=0.29$, for the substitutional surface alloy $R=0.14$, and $R=0.55$ for the compressed PdTe$_2$ layer.
Models in hcp stacking were worse than those in fcc stacking.
 
Due to the much better R-factors, only the substitutional surface alloy and the second-best honeycomb model in fcc stacking were considered for fine fits also including non-structural parameters (particularly vibrational amplitudes). 
Note that the two models differ by one additional Pd atom in the surface layer only.
The fine fit produced an excellent R-factor of $R= 0.06$ for the substitutional surface alloy model Fig.\,\ref{Fig1}(e) as final result of the LEED-IV analysis. 
At that R-factor level, there is no doubt that the correct model has been found. 
In contrast, variation of parameters of the honeycomb model did not lead to a better agreement with experiment than $R = 0.25$.

In Fig.\,\ref{Fig2}(c) and (d) we show two exemplary spectra showing the significance by which the two models can be discriminated against the experimental data.
The agreement between experiment and calculated spectra is considerably worse for the honeycomb model. 
Note that the Pendry R-factor is particularly sensitive to the energetic positions of minima and maxima due to its dependence on the logarithmic derivative of spectra \cite{Pendry1980}. 

\begin{figure}
	\centering
	\includegraphics[width=0.65\columnwidth]{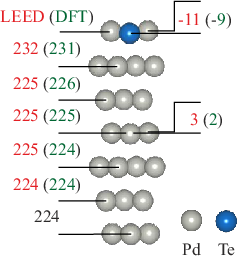}
	\caption{Structural parameters as found by LEED-IV and comparison with those found by DFT structural relaxation. On the left layer distances and on the right bucklings are given in pm. The statistical errors of the shown parameters determined by LEED amounts to less than $\pm 1\,\textrm{pm}$. The full list of varied parameters and their errors is given in the Supplement.
	}
	\label{Fig3}
\end{figure}

For the analysis we used an accumulated data base of  3.7\,keV which allowed us to fit the 17 structural and  non-structural parameters with a redundancy of $\rho =10.8$ (for the relevance of this see \cite{Kisslinger2023}). 
The structural parameters and the comparison to those obtained from the DFT structural analysis are shown in Fig.\,\ref{Fig3}. 
By virtue of the low R-factor, the error margins of the atomic z-positions in the first 4 layers are less than $\pm 1\,\textrm{pm}$, while those for the x,y positions are $\approx \pm2\,\textrm{pm}$ (see supplement for details \cite{SupMat}). 
When the DFT results are scaled from the theoretical (2.786\,\AA) to the experimental lattice parameter, the atomic positions agree perfectly within these error margins. The LEED results indicate maximal lateral shifts of 1.5\,pm from a perfect bulk crystal structure where allowed by symmetry.

We notice that among the three models with 0.33\,ML Te the DFT total energy calculations also found the substitutional model to be energetically favorable by 190\,--\,320\,meV with respect to the adsorbate or honeycomb structure in fcc (more favorable) or hcp stacking.
All detailed comparisons between experimental and calculated spectra, a list of parameter  definitions, values, and their error ranges of the model parameters are provided in the supplement \cite{SupMat}.

We have undertaken further experiments to induce the growth of a \PdDTe\ layer by deposition of 2\,ML of Te on Pd(111) at 290\,K and subsequent (careful) annealing. 
The only ordered surface structure we could obtain was the \sqrtthree-TePd$_2$ surface alloy as judged by the LEED-IV spectra. 
This indicates that after formation of the substitutional alloy, any Te in excess of 0.33\,ML simply evaporates from the surface upon annealing. 
The temperature at which this happens is tentatively determined to be 530\,K at which we start to see the development of the \sqrtthree\ superstructure evolving from a diffuse LEED image. 
This is in agreement with the observation of Liu et al. \cite{Liu2021} who deposited 4\,\AA\ Te (which most likely is more than 0.33\,ML) and observed the appearance of the well-ordered \sqrtthree\ superstructure after annealing to 470\,$^\circ$C.

\section{Conclusion}

By LEED-IV structural analysis we showed that the tellurization of Pd(111) by surface reaction with deposited Te in ultra-high vacuum is self-limited and stops at the formation of a Pd(111)-\sqrtthree-\PdTeSurf\ surface alloy.
This obviously has consequences for the interpretation of other physical or chemical properties, e.g. the XPS and HREELS data presented in \cite{Liu2021} that this system has and that should not be ascribed to \PdDTe. 

\section{Acknowledgments}
We gratefully acknowledge support from the Deutsche Forschungsgemeinschaft (DFG), project 497265814.

\section{Supplement}

Supplementary data is appended. All files that are not pdf may be opened by a simple text editor and were created by the ViPErLEED package \cite{Riva2021}. 

\begin{itemize}
	\setlength\itemsep{0pt}
	\item `EXPBEAMS.csv' containing the experimental LEED-IV data as used in the analysis.
	
	\item `POSCAR' that describes the bestfit structure in the input format used by VASP \cite{vasp3}. The entries for each atom are amended by the site number (used as reference in all other files), a site label, a layer position, an indication if the atoms are linked by symmetry and the lateral direction in which the atom may be moved due to symmetry constraints.
	
	\item `Rfactor\_plots.pdf' containing plots like Fig.\,2(c) comparing experimental and calculated spectra for all beams used in the analysis.
    
    \item `VIBROCC' that lists the vibrational amplitudes of the atoms.\\
    
	\item `Errors.pdf' with graphical representations the dependence of R-factor on parameter variation.
	
	\item `Errors.csv' summarizing the error margins in \AA\ extracted from the error curve analysis.
 
	\end{itemize}


\providecommand{\noopsort}[1]{}\providecommand{\singleletter}[1]{#1}%

\end{document}